%% file: technicalReport.tex
\documentclass[dvipsnames]{article}

\input{shorthands.tex}
\usepackage[backend=biber, style=numeric]{biblatex}
\input{packages.tex}

\usepackage{orcidlink}
\usepackage{fancyhdr}
\newgeometry{hmargin=4.7cm, vmargin=2cm}

\input{listingssettings.tex}
\input{rules/cpRules.tex}
\input{rules/epRules.tex}

\title{Verified Parameterized Choreographies\\\Large{}Technical Report}
\author{
Robert Rubbens\textsuperscript{(\Letter)}\,\orcidlink{0000-0002-5638-5945} \and
Petra van den Bos\,\orcidlink{0000-0002-9212-1525} \and
Marieke Huisman\,\orcidlink{0000-0003-4467-072X} \\\small{} Formal Methods and Tools, University of Twente, Enschede, The Netherlands \\ \small{}
\texttt{\{r.b.rubbens,p.vandenbos,m.huisman\}@utwente.nl}
}
\date{}
\begin{document}
\maketitle

\begin{abstract}
This technical report contains the full set of definitions and projection rules of the paper ``Verified Parameterized Choreographies'' by Rubbens et al.~\cite{Rubbens2025}. It also supplements the artefact~\cite{Artefact}.
\end{abstract}

\input{appendices/formalSyntax.tex}
\input{appendices/auxDefs.tex}
\input{appendices/allCpRules.tex}
\input{appendices/allEpRules.tex}

\newpage
\printbibliography
\end{document}

%% file: shorthands.tex
\usepackage{xspace}

\newcommand{\veymont}{\mbox{VeyMont}\xspace}

\newcommand{\pvl}{PVL}

\newcommand{\kwassume}{\texttt{assume}}

\newcommand{\kwtrue}{\texttt{true}}
\newcommand{\kwfalse}{\texttt{false}}

\newcommand{\kwif}{\texttt{if}}

\newcommand{\kwthis}{\texttt{this}}

\newcommand{\kwexhale}{\texttt{exhale}}

\newcommand{\kwforall}{\texttt{\textbackslash{}forall}}

\newcommand{\kwepexpr}{\texttt{\textbackslash{}endpoint}}

\newcommand{\kwchor}{\texttt{\textbackslash{}chor}}
\newcommand{\kwmsg}{\texttt{\textbackslash{}msg}}
\newcommand{\kwsender}{\texttt{\textbackslash{}sender}}
\newcommand{\kwreceiver}{\texttt{\textbackslash{}receiver}}
\newcommand{\kwrun}{\texttt{run}}

\newcommand{\kwfi}{\famidx{F}{i}}
\newcommand{\kwpar}{\texttt{par}}

\newcommand*{\chormemprojection}[2][\smallbullet]{\Braces{#1}_{#2}}

\newcommand*{\famidx}[2]{\texttt{$#1$[$#2$]}}

\newcommand*{\denote}[1]{\left\llbracket#1\right\rrbracket}
\newcommand*{\Braces}[1]{\left\lBrace#1\right\rBrace} 

\newcommand*{\braces}[1]{\texttt{\{}#1\texttt{\}}}
\newcommand*{\brackets}[1]{\texttt{[}#1\texttt{]}}
\newcommand*{\parens}[1]{\texttt{(}#1\texttt{)}}

\newenvironment{center*}[1][\smallskipamount]
  {\setlength{\topsep}{#1}\par\kern\topsep\centering}
  {\par\kern\topsep}

%% file: packages.tex
\usepackage[T1]{fontenc}

\usepackage[hidelinks]{hyperref}
\hypersetup{
	colorlinks,
	urlcolor     = blue, 
	linkcolor    = blue, 
	citecolor    = red 
}

\usepackage{color}

\usepackage[
  pass, 
]{geometry} 

\usepackage{listings}
\usepackage{chngcntr}
\usepackage{lstfiracode}
\usepackage{lstautogobble}
\usepackage{mdframed}
\mdfsetup{
  skipabove=0pt,
  skipbelow=0pt,
  leftmargin=0pt,
  rightmargin=0pt,
  innerleftmargin=0pt,
  innerrightmargin=0pt,
  innertopmargin=0pt,
  innerbottommargin=0pt
}

\usepackage{booktabs}
\usepackage{siunitx}

\usepackage{mathtools}

\usepackage{syntax}
\usepackage{amsmath}
\usepackage{amsfonts} 
\usepackage[only, llbracket, rrbracket]{stmaryrd}
\usepackage{changepage}
\usepackage{lipsum}

\usepackage{lineno}

\usepackage{mathpartir}
\usepackage{prftree}

\usepackage{adjustbox}
\usepackage{tikz}
\usetikzlibrary{fit, positioning, arrows.meta, backgrounds, shadows}
\usepackage{tcolorbox}
\tcbuselibrary{skins}
\usepackage[disable]{todonotes}

\usepackage[misc]{ifsym}
\usepackage{fontawesome}

\usepackage{subcaption}
\usepackage{float}

\usepackage[inline, shortlabels]{enumitem}

\makeatletter
\@ifpackageloaded{stix}{%
}{%
  \DeclareFontEncoding{LS2}{}{\noaccents@}
  \DeclareFontSubstitution{LS2}{stix}{m}{n}
  \DeclareSymbolFont{stix@largesymbols}{LS2}{stixex}{m}{n}
  \SetSymbolFont{stix@largesymbols}{bold}{LS2}{stixex}{b}{n}
  \DeclareMathDelimiter{\lBrace}{\mathopen} {stix@largesymbols}{"E8}%
                                            {stix@largesymbols}{"0E}
  \DeclareMathDelimiter{\rBrace}{\mathclose}{stix@largesymbols}{"E9}%
                                            {stix@largesymbols}{"0F}
}
\makeatother

\usepackage[
  capitalise, 
  nameinlink, 
]{cleveref}

\crefname{line}{line}{lines}
\crefname{page}{page}{pages}

\newcounter{inferrule}
\let\oldinferrule\inferrule
\makeatletter
\newcommand*{\myspecialwrite}[2]{%
\immediate\write\@auxout{\unexpanded{\global\@namedef{#1}{#2}}}%
}
\renewcommand*{\inferrule}[3][]{\oldinferrule[#1]{#2}{#3}}
\newcommand*{\labelinferrule}[4]{\oldinferrule[\refstepcounter{inferrule}#1]{#3}{#4}
\label{inferrulelabel@#1:#2}%
\expandafter\myspecialwrite\expandafter{\expanded{inferrulename@#1}}{#1}%
}

\crefformat{inferrule}{rule #2#1#3}

\newcommand*{\labelinferrulename}[1]{%
\ifcsname inferrulename@#1\endcsname {\mbox{\@nameuse{inferrulename@#1}}}\else{%
  \GenericWarning{[nonexistentrulename]}{Rule label "#1" does not exist}{}{}}\fi}
\newcommand*{\prettyrule}[1]{\textsc{\labelinferrulename{#1}}}
\newcommand*{\refrule}[2][paper]{{\hypersetup{allcolors=black}{\hyperref[inferrulelabel@#2:#1]{\prettyrule{#2}}}}}
\makeatother



%% file: listingssettings.tex
\colorlet{pvlNumberColor}{PineGreen}
\definecolor{contractPurple}{HTML}{9e30c7}
\lstdefinelanguage{PVL}{
  morekeywords={
    communicate,
    if,
    else,
    while,
    with,
    choreography,
    communication,
    pure,
    endpoint,
    endpoints,
    run,
    par,
    class,
    lock,
    unlock,
    \\in,
    inline,
  },
  morekeywords = [2]{
    int,
    boolean,
    seq,
    resource,
    void,
  },
  morekeywords = [3]{
    fold,
    unfold,
    \\forall,
    \\exists,
    context,
    context\_everywhere,
    Perm,
    lock\_invariant,
    requires,
    ensures,
    channel\_invariant,
    loop\_invariant, invariant,
    assert,
    exhale,
    inhale,
    \\old, 
    \\unfolding,
    \\sender,
    \\receiver,
    \\msg,
    \\endpoint,
    \\endpoints,
    \\chor,
  },
  morekeywords = [4]{
    true, false
  },
  alsoletter={\\},
  literate=%
   *{0}{{{\color{pvlNumberColor}0}}}1%
    {1}{{{\color{pvlNumberColor}1}}}1%
    {2}{{{\color{pvlNumberColor}2}}}1%
    {3}{{{\color{pvlNumberColor}3}}}1%
    {4}{{{\color{pvlNumberColor}4}}}1%
    {5}{{{\color{pvlNumberColor}5}}}1%
    {6}{{{\color{pvlNumberColor}6}}}1%
    {7}{{{\color{pvlNumberColor}7}}}1%
    {8}{{{\color{pvlNumberColor}8}}}1%
    {9}{{{\color{pvlNumberColor}9}}}1%
    ,%
  moredelim=[is][\color{black}]{[*}{*]},
  morecomment=[l]{//}, 
  morecomment=[s]{/*}{*/}, 
  morestring=[b]" 
} %

\lstdefinestyle{java}{
  escapeinside={(*@}{@*)},
	language=Java,
  style=FiraCodeStyle,
	numbers=left,
	numberstyle=\footnotesize\color{darkgray},
	tabsize=2,
	columns=fixed,
	showstringspaces=false,
	showtabs=false,
	keepspaces,
	commentstyle=\color{MidnightBlue}\ttfamily,
	keywordstyle=\color{blue}\ttfamily,
	basicstyle=\ttfamily\footnotesize,
  breaklines, breakatwhitespace
}
\lstdefinestyle{pvl}{
  escapeinside={(*@}{@*)},
  style=FiraCodeStyle,
  language=PVL,
	numbers=left,
	numberstyle=\scriptsize\color{darkgray},
	tabsize=2,
	columns=fixed,
	showstringspaces=false,
	showtabs=false,
	keepspaces,
	commentstyle=\color{gray}\ttfamily,
	keywordstyle=\color{blue}\ttfamily,
  keywordstyle=[2]\color{blue}\ttfamily,
  keywordstyle=[3]\color{contractPurple}\ttfamily,
  keywordstyle=[4]\color{blue}\ttfamily,
	basicstyle=\ttfamily\scriptsize,
  breaklines, breakatwhitespace
}

%% file: rules/cpRules.tex
\newcommand*{\RuleCpExpr}[1][paper]{
\labelinferrule{CpExpr}{#1}{}{
\Braces{\texttt{(\kwepexpr{} $r$; $E$)}} = \Braces{E}_r
}
}

\newcommand*{\RuleCpExprSkip}[1][paper]{
\labelinferrule{CpExprSkip}{#1}{}{
{\begin{array}{r}
\Braces{\texttt{(\kwepexpr{} $\alpha$; $E$)}}_r = \texttt{true} \\
\textrm{if $\textsf{sort}(\alpha) \neq \textsf{sort}(r)$}
\end{array}}
} 
}

\newcommand*{\RuleCpAssign}[1][paper]{
\labelinferrule{CpAssign}{#1}{}{
\Braces{\texttt{$r$: $H_{loc}$ := $H_v$;}} = \texttt{$\Braces{H_{loc}}_r$ = $\Braces{H_v}_r$;}
}
}

\newcommand*{\RuleCpIf}[1][paper]{
\labelinferrule{CpIf}{#1}{}{
{
\Braces{\texttt{if ($H$) $S_{\kwtrue{}}$ $S_{\kwfalse{}}$}} = \begin{array}{l}
\texttt{\{~assert \textsf{unanimous}$(H)$;} \\
\texttt{~~if ($\Braces{H}$) $\Braces{S_{\kwtrue{}}}$ $\Braces{S_{\kwfalse{}}}$ \}}
\end{array}
}
} 
}

\newcommand*{\RuleCpWhile}[1][paper]{
\labelinferrule{CpWhile}{#1}{}{
{
\begin{array}{l}
\Braces{\texttt{loop\_invariant $R_{inv}$; while ($H_{cond}$) $S$}} = \\
\hspace{0.45cm}\texttt{loop\_invariant $\textsf{unanimous}(R_{inv})$ \&\& $\Braces{R_{inv}}$;} \\
\hspace{0.45cm}\texttt{while ($\Braces{H_{cond}}$) $\Braces{S}$}
\end{array}
}
}
}

\newcommand*{\RuleCpMethodCall}[1][paper]{
\labelinferrule{CpMethodCall}{#1}{}{
{
\begin{array}{l}%
\Braces{
\texttt{endpoint $r$: $H$.$m$();}
} = \chormemprojection[H]{r}.\chormemprojection[m\texttt{()}]{r};
\end{array}
}
}
}

\newcommand*{\RuleCpComm}[1][paper]{
\labelinferrule{CpComm}{#1}{}{
{
\begin{array}{l}
\Braces{
\begin{array}{l}
\texttt{channel\_invariant $R_I(\kwmsg{}, \; \kwsender{}, \; \kwreceiver{})$;} \\
\texttt{communicate $r$: $H_{msg}$ -> $p$: $H_{dst}$;}
\end{array}
} = \\[7.5pt]
\hspace{0.6cm}\texttt{\{ $T$ $v$ = $\Braces{H_{msg}}_{r}$;} \\
\hspace{0.6cm}\texttt{~~exhale $\Braces{R_I(v, \; r, \; p)}_{r}$;} \\
\hspace{0.6cm}\texttt{~~inhale $\Braces{R_I(v, \; r, \; p)}_{p}$;} \\
\hspace{0.6cm}\texttt{~~$\Braces{H_{dst}}_p$ = v; \}}
\end{array}
}
}
}

\newcommand*{\RuleCpExprRange}[1][paper]{
\labelinferrule{CpExprRange}{#1}{}{
{
\begin{array}{l}
\Braces{\texttt{(\kwepexpr{} $F$[$i$ := $E_l$ .. $E_h$]; $E$)}} = \\
\hspace{0.3cm}\texttt{(\kwforall{} int $i$ = $E_l$ .. $E_h$; $\Braces{E}_{\kwfi}$)}
\end{array}
}
}
}

\newcommand*{\RuleCpExprIndex}[1][paper]{
\labelinferrule{CpExprIndex}{#1}{}{
{
\begin{array}{l}
\Braces{\texttt{(\kwepexpr{} $F$[$j$ := $E_l$ .. $E_h$]; $E(j)$)}}_\kwfi{} = \\
\hspace{0.6cm}\texttt{$E_l$ <= $i$ \&\& $i$ < $E_h$ ==> $\Braces{E(i)}_\kwfi{}$}
\end{array}
}
}
}

\newcommand*{\RuleCpMethodCallRange}[1][paper]{
\labelinferrule{CpMethodCallRange}{#1}{}{
{
\begin{array}{l}%
\Braces{
\texttt{endpoint $F$[$i$ := $E_l$ .. $E_h$]: $F$[$i$].$m$();}
} = \\
\hspace{0.6cm}\texttt{par (int $i$ = $E_l$ .. $E_h$)} \\
\hspace{0.9cm}\texttt{requires $\Braces{\textsf{pre($m$, \kwfi{})}}_{\kwfi{}}$;} \\
\hspace{0.9cm}\texttt{ensures $\Braces{\textsf{post($m$, \kwfi{})}}_{\kwfi{}}$;} \\
\hspace{0.6cm}\texttt{\{ $\Braces{\texttt{endpoint \kwfi{}: \kwfi{}.$m$();}}$ \}}
\end{array}
}
}
}

\newcommand*{\RuleCpCommRange}[1][paper]{
\labelinferrule{CpCommRange}{#1}{}{
{
\begin{array}{l}
\Braces{
\begin{array}{l}
\texttt{channel\_invariant $R_I(\kwmsg{}, \kwsender{}, \kwreceiver{})$;} \\
\texttt{communicate $F$[$i$ := $E_l$ .. $E_h$]: $F$[$i$].$f$ -> $G$[$d(i)$]: $G$[$d(i)$].$g$;}
\end{array}
} = \\[7.5pt]
\hspace{0.6cm}\texttt{assert (\kwforall{} int i, j = $E_l$ .. $E_h$; $d(\texttt{i})$ == $d(\texttt{j})$ ==> i == j);}\\  
\hspace{0.6cm}\texttt{par (int $i$ = $E_l$ .. $E_h$)}\\
\hspace{0.9cm}\texttt{context $\Braces{\texttt{Perm($F$[$i$].$f$, $\epsilon$)}}_{\famidx{F}{i}}$ ** $\Braces{\texttt{Perm(\famidx{G}{d(i)}.$g$, 1)}}_\famidx{G}{d(i)}$;} \\
\hspace{0.9cm}\texttt{requires $\Braces{R_I(\texttt{$F$[$i$].$f$}, \; \kwfi{}, \; \famidx{G}{d(i)})}_{\famidx{F}{i}}$;} \\
\hspace{0.9cm}\texttt{ensures $\Braces{R_I(\texttt{\famidx{G}{d(i)}.$g$}, \; \kwfi{}, \; \famidx{G}{d(i)})}_\famidx{G}{d(i)}$;} \\
\hspace{0.6cm}\texttt{\{~$T$ $v$ = $\Braces{\texttt{$F$[$i$].$f$}}_\kwfi$;} \\
\hspace{0.6cm}\texttt{~~exhale $\Braces{R_I(v, \; \texttt{$F$[$i$]}, \; \texttt{$G$[$d(i)$]})}_\kwfi$;} \\
\hspace{0.6cm}\texttt{~~inhale $\Braces{R_I(v, \; \kwfi{}, \; \famidx{G}{d(i)})}_\famidx{G}{d(i)}$;} \\
\hspace{0.6cm}\texttt{~~$\Braces{\texttt{$\famidx{G}{d(i)}$.$g$}}_\famidx{G}{d(i)}$ = v; \}}
\end{array}
}
}
}

%% file: rules/epRules.tex
\newcommand*{\RuleEpAssign}[1][paper]{
\labelinferrule{EpAssign}{#1}{}{
\denote{\texttt{$e$: $H_{loc}$ := $H_v$;}}_e = \texttt{$H_{loc}$ = $H_v$;}
}
}

\newcommand*{\RuleEpAssignSkip}[1][paper]{
\labelinferrule{EpAssignSkip}{#1}{}{
\denote{\texttt{$e$: $H_{loc}$ := $H_v$;}}_r = \texttt{\braces{}}
}
}

\newcommand*{\RuleEpSend}[1][paper]{
\labelinferrule{EpSend}{#1}{}{
{
\begin{array}{l}
\denote{\texttt{$L$: communicate $a$: $H_{msg}$ -> $b$: $H_{dst}$;}}_a^{send} = \\
\hspace{0.3cm} \texttt{$\denote{L}_a$.writeValue($H_{msg}$) with \{} \\
\hspace{0.6cm} \texttt{sender = $a$; receiver = $b$; \};}
\end{array}
}
}
}

\newcommand*{\RuleEpReceive}[1][paper]{
\labelinferrule{EpReceive}{#1}{}{
{
\begin{array}{l}
\denote{\texttt{$L$: communicate $a$: $H_{msg}$ -> $b$: $H_{dst}$;}}_b^{receive} = \\
\hspace{0.3cm} \texttt{$H_{dst}$ = $\denote{L}_b$.readValue() with \{} \\
\hspace{0.6cm} \texttt{sender = $a$; receiver = $b$; \};}
\end{array}
}
}
}

\newcommand*{\RuleEpComm}[1][paper]{
\labelinferrule{EpComm}{#1}{}{
{\begin{array}{l}
\denote{
\begin{array}{ll}
\texttt{$L$:} & \texttt{channel\_invariant $R_{inv}$;} \\
              & \texttt{communicate $\alpha$: $H_{msg}$ -> $\beta$: $H_{dest}$;} \\
\end{array}
}_r = \\[10pt]
\hspace{0.7cm}\denote{\texttt{$L$: communicate $\alpha$: $H_{msg}$ -> $\beta$: $H_{dst}$;} \;}^{send}_r \texttt{;} \\
\hspace{0.7cm}\denote{\texttt{$L$: communicate $\alpha$: $H_{msg}$ -> $\beta$: $H_{dst}$;} \;}^{receive}_r \texttt{;} 
\end{array}}
}
}

\newcommand*{\RuleEpCommSkip}[1][paper]{
\labelinferrule{EpCommSkip}{#1}{}{
{\begin{array}{r}
\denote{
\texttt{$L$: communicate $\alpha$: $H_{msg}$ -> $\beta$: $H_{dst}$;}
}^-_r = \texttt{\{\}} \\
\textrm{if $\textsf{sort}(\alpha)$, $\textsf{sort}(\beta)$, $\textsf{sort}(r)$ pairwise distinct}
\end{array}}
}
}

\newcommand*{\RuleEpExpr}[1][paper]{
\labelinferrule{EpExpr}{#1}{}{
\denote{\texttt{(\textbackslash{}endpoint $e$; $E$)}}_e = E
}
}

\newcommand*{\RuleEpExprSkip}[1][paper]{
\labelinferrule{EpExprSkip}{#1}{}{
{
\begin{array}{r}
\denote{\texttt{(\textbackslash{}endpoint $\alpha$; $E$)}}_r = \texttt{true} \\
\textrm{if $\textsf{sort}(\alpha) \neq \textsf{sort}(r)$}
\end{array}
}
}
}

\newcommand*{\RuleEpExprIndex}[1][paper]{
\labelinferrule{EpExprIndex}{#1}{}{
{ 
\begin{array}{l}
\denote{
\texttt{(\textbackslash{}endpoint \famidx{F}{E_i}; $E$)}
}_\kwfi = \texttt{$i$ == $E_i$ ==> $E$}
\end{array}
}
}
}

\newcommand*{\RuleEpRange}[1][paper]{
\labelinferrule{EpRange}{#1}{}{
{
\denote{
\texttt{(\textbackslash{}endpoint $F$[$j$ := $E_l$ .. $E_h$]; $E$)}
}_\kwfi = \texttt{$E_l$ <= $i$ \&\& $i$ < $E_h$ ==> $E$}
}
}
}

\newcommand*{\RuleEpAnd}[1][paper]{
\labelinferrule{EpAnd}{#1}{}{
\denote{
\texttt{$E_1$ \&\& $E_2$}
}_r = \texttt{$\denote{E_1}_r$ \&\& $\denote{E_2}_r$}
}
}

\newcommand*{\RuleEpChor}[1][paper]{
\labelinferrule{EpChor}{#1}{}{
\denote{
\texttt{(\textbackslash{}chor $E$)}
}_r = \texttt{true}
}
}

\newcommand*{\RuleEpIf}[1][paper]{
\labelinferrule{EpIf}{#1}{}{
\denote{\texttt{if ($H$) $S_{true}$ $S_{false}$}}_r = \texttt{if ($\denote{H}_r$) $\denote{S_{true}}_r$ $\denote{S_{false}}_r$}
}
}

\newcommand*{\RuleEpWhile}[1][paper]{
\labelinferrule{EpWhile}{#1}{}{
{
\begin{array}{l}
\denote{\texttt{loop\_invariant $R_{i}$; while ($H$) $S$}}_r = \\
\hspace{0.3cm}\texttt{loop\_invariant $\denote{R_{i}}_r$; while ($\denote{H}_r$) $\denote{S}_r$}
\end{array}
}
}
}

\newcommand*{\RuleEpIndexSend}[1][paper]{
\labelinferrule{EpIndexSend}{#1}{}{
{\begin{array}{l}
\denote{\texttt{$L$: communicate $F$[$E_j$]: $H_{msg}$ -> $r$: $H_{dst}$;}}_\kwfi^{send} = \\
\hspace{0.3cm} \texttt{if ($i$ == $E_j$) \{}\\
\hspace{0.6cm} \texttt{$\denote{L}_\kwfi$.writeValue($\denote{H_{msg}}_\kwfi$) with \{} \\
\hspace{0.9cm} \texttt{sender = $F$[$i$]; receiver = $r$; \}; \}}
\end{array}}
}
}

\newcommand*{\RuleEpIndexReceive}[1][paper]{
\labelinferrule{EpIndexReceive}{#1}{}{
{\begin{array}{l}
\denote{\texttt{$L$: communicate $r$: $H_{msg}$ -> $F$[$E_j$]: $H_{dst}$}}_\kwfi^{receive} = \\
\hspace{0.3cm} \texttt{if ($i$ == $E_j$) \{}\\
\hspace{0.6cm} \texttt{$\denote{H_{dst}}_\kwfi$ = $\llbracket L \rrbracket_\kwfi$.readValue() with \{} \\
\hspace{0.9cm} \texttt{sender = $r$; receiver = $F$[$i$]; \}; \}}
\end{array}}
}
}

\newcommand*{\RuleEpRangeSend}[1][paper]{
\labelinferrule{EpRangeSend}{#1}{}{
{\begin{array}{l}
\denote{
\texttt{$L$: communicate $F$[$j$: $E_l$ .. $E_h$].$f$ -> $G$[$d(j)$].$g$}
}_\kwfi^{send} = \\
\hspace{0.3cm} \texttt{if ($E_l$ <= $i$ \&\& $i$ < $E_h$) \{}\\
\hspace{0.6cm} \texttt{$\denote{L}_{\famidx{F}{i}}$[$i$].writeValue($F$[$i$].$f$) with \{} \\
\hspace{0.9cm} \texttt{sender = $F$[$i$]; receiver = $G$[$d(i)$]; \}; \}}
\end{array}}
}
}

\newcommand*{\RuleEpRangeReceive}[1][paper]{
\labelinferrule{EpRangeReceive}{#1}{}{
{\begin{array}{l}
\denote{
\texttt{$L$: communicate $F$[$j$: $E_l$ .. $E_h$].$f$ -> $G$[$d(j)$].$g$}
}_{\famidx{G}{i}}^{receive} = \\
\hspace{0.3cm} \texttt{if ($E_l$ <= $d^{-1}(i)$ \&\& $d^{-1}(i)$ < $E_h$) \{}\\
\hspace{0.6cm} \texttt{\famidx{G}{i}.$g$ = $\denote{L}_{\famidx{G}{i}}$[$d^{-1}(i)$].readValue() with \{} \\
\hspace{0.9cm} \texttt{sender = $F$[$d^{-1}(i)$]; receiver = $G$[$i$]; \}; \}}
\end{array}}
}
}

%% file: appendices/formalSyntax.tex
\section{Complete formal syntax} \label{app:syntax}

The figure below is the formal syntax of PVL programs supported by VeyMont, including the complete OOP fragment of PVL.

\begin{figure}[t]
\begin{adjustwidth}{-0.15\paperwidth}{-0.15\paperwidth}
\begin{minipage}{0.5\textwidth}
\begin{align*}
x, y, z &::= \textrm{field,} \quad v, u, w ::= \textrm{variable,} \quad m ::= \textrm{method,}\\
f &::= \textrm{function,} \quad C ::= \textrm{class,} \quad P ::= \textrm{predicate,} \\
T &::= \texttt{int} \mid \texttt{boolean} \mid \texttt{seq<}T\texttt{>} \mid C \mid ... \\
K &::= \texttt{requires $E$; ensures $E$;} \qquad  \\
K_H &::= \textrm{$K$ with $R, H$} \qquad K_R ::= \textrm{$K$ with $R, R$} \\
\textrm{\textbf{prog}} &::= \overline{\textrm{\textbf{decl}}} \\
\textrm{\textbf{decl}} &::= \texttt{class} \; C \; \texttt{\{} \; \overline{D_{cls}} \; \texttt{\}} 
\mid \texttt{resource} \; P\parens{\overline{T \; v}} \; = \; E\texttt{;} \\
& \mid \texttt{$K$ pure} \; T \; f \parens{\overline{T \; v}} \; \texttt{=} \; E\texttt{;} \mid \texttt{$K_H$ pure} \; T \; f_H \parens{\overline{T \; v}} \; \texttt{=} \; H\texttt{;} \\
& \mid \textrm{\textbf{chor}} \\
D_{cls} &::= \; T \; x\texttt{;} \mid K_R \; T \; m \texttt{(} \overline{T \; v} \texttt{)} \; S_m \\
E &::= v \mid r \mid F \mid E \; \texttt{+} \; E \; | \; E \; \texttt{\&\&} \; E \mid E\texttt{.}f\parens{\overline{E}} \mid ... \\
H &::= \textrm{$E$ extended with: $\texttt{$E_h$.$x$} \mid \texttt{$E_h$.$f_h$($\overline{E_h}$)}$} \mid \texttt{$F$[$E_h$]} \mid \texttt{this} \\
R &::= H \mid \texttt{Perm($H$.$x$, $H$)} \mid \texttt{$R$ ** $R$} \mid \texttt{$P$($\overline{H}$)} \\
& \mid \texttt{$H$ ==> $R$} \mid \textrm{QP} \\
S_{m} &::= \texttt{assert} \; H \texttt{;} \mid H \; \texttt{=} \; H\texttt{;} \mid \texttt{$H$.$m$($\overline{H}$)} \\
& \mid \texttt{inhale $R$;} \mid \texttt{exhale $R$;} \mid \texttt{if} \; \parens{H} \; S_m \; S_m \\
& \mid \texttt{loop\_invariant} \; R\texttt{;} \; \texttt{while} \; \parens{H} \; S_m \mid ...\\
& \mid K_R \; \texttt{par} \; \parens{T \; v \; = \; H \; \texttt{..} \; H} \; S_m 
\end{align*}
\end{minipage}%
\hfill
\begin{minipage}{0.5\textwidth}
\begin{align*}
e, a, b &::= \textrm{endpoint} \quad F, G ::= \textrm{endpoint family,} \\
\textrm{\textbf{chor}} &::= \texttt{$K_R$ choreography} \parens{\overline{T \; v}} \braces{ \; \overline{D_{\texttt{chor}}} \;} \\
D_{chor} &::= \texttt{endpoint} \; e \; \texttt{=} \; C \parens{\overline{H}} ; \\
& \mid \texttt{endpoint} \; F \brackets{v \; \texttt{:=} \; 0 \; \texttt{..} \; H} \; = \; C \parens{\overline{E}} ; \\
& \mid K_R \; \texttt{run} \; \braces{\; S_{\texttt{chor}} \;}  \\
S_{chor} &::= \texttt{if} \; \parens{H_{chor}} \; S_{chor} \; S_{chor} \mid \texttt{assert} \; R_{chor}\texttt{;} \\
& \mid \texttt{loop\_invariant $R_{chor}$; while ($H_{chor}$) $S_{chor}$} \\
& \mid \texttt{endpoint} \; \alpha\texttt{:} \; S_{ep} \\
& \mid \texttt{channel\_invariant} \; R_{chan}\texttt{;} \; \texttt{communicate} \; \alpha\texttt{:} \; H \; \texttt{->} \; \alpha\texttt{:} \; H\texttt{;} \\
S_{ep} &::= H\texttt{.}m\parens{\overline{H}} \mid \texttt{$H$ := $H$;} \\
H_{chor} &::= \parens{\kwepexpr{} \; \alpha\texttt{;} \; H} \mid H_{chor} \; \texttt{\&\&} \; H_{chor} \\ 
R_{chor} &::= \parens{\kwepexpr{} \; \alpha\texttt{;} \; R} \mid R_{chor} \; \texttt{\&\&} \; R_{chor} \mid \parens{\kwchor{} \; H} \\ 
R_{chan} &::= \textrm{$R$ extended with: $\kwmsg{} \mid \kwsender{} \mid \kwreceiver{}$} \\
r, p &::= e \mid F\brackets{E} \qquad \alpha, \beta ::= r \mid F\brackets{v \; \texttt{:=} \; E \; \texttt{..} \; E}
\end{align*}
\end{minipage}%
\end{adjustwidth}
\caption{PVL syntax. Left: OOP fragment, right: choreographic fragment.}\label{app:fig:syntax}
\end{figure}

\paragraph{Core PVL} The syntax of PVL is shown in \cref{app:fig:syntax}, of which the left is the syntax for core OOP PVL. PVL has classes, methods, fields, and supports several built-in types, such as integers (\texttt{int}), booleans, and sequences (e.g. \texttt{seq<int>}). It supports standard statements such as \texttt{while}, \texttt{if} and variable assignment, and standard expressions such as boolean logic and arithmetic. It also supports verification primitives such as contracts with pre- and postconditions, and assertions and ownership through permission annotations. The main primitive for concurrency in \pvl{} is the \texttt{par} block. When a main thread reaches a \texttt{par} block, $N$ subthreads are spawned to execute the body of the \texttt{par} block in parallel. The main thread waits until all subthreads are finished, and then continues with the remainder of the program.

We want to highlight how expressions are defined in \cref{app:fig:syntax}. Pure expressions $E$ only depend on local variables and immutable value constructors, such as sequences and sets. Heap-dependent expressions $H$ are a superset of $E$, that can also refer to fields of objects. Resource expressions ($R$) are a superset of $H$, and include permissions using the \texttt{Perm} keyword, as well as the separating conjunction operator \texttt{**} to compose resources. These different kinds of expressions give rise to variations of several other nodes: pure contracts ($K$), contracts that only inspect the heap and not modify it ($K_H$) and contracts that require and return resources ($K_R$), pure functions ($f$) and functions that read the heap ($f_H$). 

\paragraph{Choreography DSL in PVL} On the right of \cref{app:fig:syntax} is the choreographic fragment of PVL. A choreography has zero or more parameters, and defines one or more endpoint or endpoint families. Endpoint $e$ are singular endpoints, defined to have a name and instructions on how it should be instantiated with a given constructor. Endpoint families extend this notion with an extra size parameter that indicates the size of the endpoint family at runtime. Where a singular endpoint is represented at run-time with an instance of the given class $C$, an endpoint family is represented with an immutable sequence of such instances. Finally, a choreography also contains a \kwrun{} declaration, which is the main body of the choreography and contains a sequence of choreographic statements.

There are a two ways to refer to endpoints. First, to refer to singular endpoints there is the notation $r$, which refers to either an endpoint $e$ or a member of an endpoint family $F$ at index $E$. Second, there is the notation $\alpha$ for endpoint targets in general, which extends singular endpoints with ranges of endpoint families. This can be used to state that e.g. a statement must be executed by a subrange of an endpoint family.

The syntax for choreographic statements $S_{chor}$ partially overlaps with regular PVL statements $S_m$, but they cannot be used interchangeably. Specifically because e.g. the choreographic \kwif{} requires its condition to be one or more endpoint expressions \texttt{(\kwepexpr{} $\alpha$; $H$)} combined with \texttt{\&\&}, whereas the regular PVL \kwif{} requires a plain heap expression $H$. The semantics is similar to the semantics of each statement in PVL, except that each endpoint only executes those staments that are relevant to the endpoint. For example, the choreographic statement \texttt{endpoint $a$: $e$.$x$ := 3} will be executed by $a$, but skipped by $b$. Composite statements are transparent with regards to this; if an endpoint does not occur within a composite choreographic statement, it skips it.

A communication statement is parameterized when the $\alpha$ notation is used to denote a range, such as \texttt{$F$[v := 0 .. N]}. When both alphas of the communicate indicate a singular endpoint, it is just a regular non-parameterized communicate (though even in the non-parameterized communicate, a member of an endpoint family might participate through endpoint family indexing, such as \texttt{$F$[5]}).

The user can also declare a channel invariant on a communicate statement, which specifies an invariant over values sent over that channel. This invariant must be proven over the values sent, and may be assumed over the values received. Within this invariant, the keywords \kwsender{}, \kwreceiver{} and \kwmsg{} must be used to refer symbolically to the respective concepts.

Choreographic expressions are adapted to allow the same projectability that is natural for statements. The primitive here is the endpoint expression, written as \texttt{(\kwepexpr{} $\alpha$; $H$)}, and also for $R$. This expression indicates that $H$ is only relevant for the endpoint target $\alpha$, and must be ignored by other endpoints. This means that any endpoint covered by $\alpha$ will evaluate the expression, while endpoints not covered by $\alpha$ will simply continue as if the expression evaluated to \kwtrue{}. This is sound, because endpoint expressions may only occur positively, and because VeyMont checks branch unanimity~\cite{VandenBos2023}.

The \kwchor{} expression is used to indicate that an expression should only be included in the choreographic projection, and that is should not be included in the endpoint projection. In addition, within a \kwchor{} expression, permissions from all endpoints can freely be mixed. In spirit, \kwchor{} is similar to \kwassume{}, in that it is used to ``debug'' non-verifying programs, and that any use of \kwchor{} should include an explanation of why it is needed, or otherwise removed. For more information on endpoint expressions, we refer the reader to \cite{Rubbens2024}.

%% file: appendices/auxDefs.tex
\section{Auxiliary definitions}

\begin{figure}[t]
\begin{adjustwidth}{-0.15\paperwidth}{-0.15\paperwidth}
\begin{mathpar}
\inferrule[Sort]{}{
{
\begin{array}{cl}
\textsf{sort}(e) & =  e \\
\textsf{sort}(\texttt{$F$[$i$]}) & = F \\
\textsf{sort}(\texttt{$F$[$i$ := $E_{low}$ .. $E_{high}$]}) & = F
\end{array}
}
}
\end{mathpar}
\end{adjustwidth}
\caption{Definition of \textsf{sort} to approximate inequality between instances of $\alpha$}\label{fig:aux}
\end{figure}

\Cref{fig:aux} shows the auxiliary definition \textsf{sort}.
The function \textsf{sort} approximates inequalities on $\alpha$. This is useful for checking if it is possible for two $\alpha$ notations to be equal. For example, if $\textsf{sort}(\texttt{$F$[$i$]}) \neq \textsf{sort}(\texttt{$G$[$j$]})$, then \texttt{$F$[$i$]} and \texttt{$G$[$j$]} are also distinct.

We define functions $\textsf{pre}(m, E)$ and $\textsf{post}(m, E)$ axiomatically to return the pre-/postcondition of $m$. In addition, they replace any occurrence of \kwthis{} in the return value with $E$. If these functions are given a class $C$, they return the pre-/postcondition of the constructor of the class.

%% file: appendices/allCpRules.tex
\section{Choreographic Projection Rules}\label{app:cp}

\begin{figure}[ht!]
\centering
\begin{adjustwidth}{-0.17\paperwidth}{-0.17\paperwidth}
\begin{mathpar}
\RuleCpExpr[tr]{} \and 
\RuleCpExprSkip[tr]{} \and
\RuleCpAssign[tr]{} \and
\RuleCpIf[tr]{} \and
\RuleCpWhile[tr]{} \and
\RuleCpMethodCall[tr]{} \and
\RuleCpComm[tr]{} \and
\RuleCpExprRange[tr]{} \and 
\RuleCpExprIndex[tr]{} \and
\RuleCpMethodCallRange[tr]{} \and
\RuleCpCommRange[tr]{}
\end{mathpar}
\end{adjustwidth}
\caption{All choreographic projection rules}
\end{figure}

We will now discuss the transformation rules for the choreographic projection.

The transformation rules distinguish between \begin{enumerate*}[1.]\item singular endpoints using $e$, \item  singular or indexed endpoint families using $r$, and \item endpoints or endpoint family ranges using $\alpha$\end{enumerate*}. E.g. rule \refrule[tr]{CpExpr} applies only to singular endpoints, such that another rule (in this case rule \refrule[tr]{CpExprRange}) is necessary to handle the parameterized case. In contrast, e.g. rule \refrule[tr]{CpExprSkip} uses $\alpha$, and hence works for both endpoint family ranges as well as singular endpoints.

Rule \refrule[tr]{CpExpr} enables confined memory mode~\cite{Rubbens2024} to make sure $E$ is evaluated using only memory of $r$.

Rule \textsc{CpExprSkip} skips an expression by transforming it to \kwtrue{} if it is not relevant for the current target for confinement. This is safe, because the side condition $\textsf{sort}(\alpha) \neq \textsf{sort}(r)$ guarantees that the expression is not relevant to $r$.

Rule \refrule[tr]{CpAssign} uses confined memory mode to ensure the assignment is executed on the memory of $r$. For this rule there is no parameterized version. This is because it is difficult to automatically infer the footprint of the expressions $E_{loc}$ and $E_v$ in a parameterized context. If required, the user can work around this by defining a method on an endpoint that only writes to a field, and call this using the rule for parameterized method invocation, discussed later.

The rules \refrule[tr]{CpIf} and \refrule[tr]{CpWhile} forward the choreographic projection to their subparts, while also adding deadlock freedom checks~\cite{VandenBos2023}. As there is no endpoint context on these statements, no confinement is necessary.

Rule \refrule[tr]{CpMethodCall} evaluates the target of the method in confined memory mode. On then target, it calls a version of the method $m$ adapted to the stratified permissions memory model~\cite{Rubbens2024}.

Rule \refrule[tr]{CpComm} encodes a communication from endpoint $r$ to endpoint $p$. First, the message value is computed, confined to the memory of $r$. Then, the channel invariant is removed from the state of $r$ using the \kwexhale{} statement. Note that the channel may contain the placeholder expressions \kwmsg, \kwsender, \kwreceiver, in this case referring to the value of $H_{msg}$, $r$ and $p$ respectively. The projection instantiates these placeholders with their concrete values by passing $v$, $r$ and $p$ as arguments to the channel invariant $R_I$. Then, the invariant is added to the state of $p$, after which finally the value is written to the destination location.

Rule \refrule[tr]{CpExprRange} evaluates an expression for all endpoints in an endpoint family symbolically by replacing the \kwepexpr{} keyword with \kwforall{}. This is sound, as \kwepexpr{} expressions can only occur in a positive positions: $H_{chor}$ is essentially a list of \kwepexpr{} expressions combined with \texttt{\&\&}. This is at the logical level equivalent to using \kwforall{}.

Rule \refrule[tr]{CpExprIndex} shows how to project an endpoint expression with a range in confined mode: an implication is prepended to the expression $E$ that ensures $E$ is only evaluated if the confinement target index $j$ is within the bounds of the endpoint expression range, $E_{low}$ and $E_{high}$. Effectively, we intersect the range specified by the endpoint expression with the confinement target. This rule is necessary when using the confined memory mode for branch unanimity (i.e. the \textsf{unanimous} function).

Rule \refrule[tr]{CpMethodCallRange} transforms a method call on a range of endpoints into a par block that executes the method calls indepently and in parallel. This is essential: if the par block can be proven correct, this means the method calls can safely be executed independly and in parallel, which means splitting this method call up using the endpoint projection is safe. The syntax for the object on which the method is called is restricted: instead of a general expression $H$ we allow only an indexed family. This ensures the required annotations for the \kwpar{} block can be automatically generated, as it keeps the footprint predictable and exact.

Rule \refrule[tr]{CpCommRange} does something similar as rule \refrule[tr]{CpMethodCallRange}, except for two things. First, the injectivity is checked by adding an assert and a quantifier encoding the injectivity property over the expression $d$. Second, by modelling the actual message exchange within the par block. This message exchange works as follows: \begin{enumerate*}\item evaluate the message in the context of \kwfi{}, \item remove the channel invariant from the state of \kwfi{}, \item add the state to \famidx{G}{d(i)}, \item  assign the message to the destination location, allowing only memory to be used of the receiving party\end{enumerate*}. Similar to rule \refrule[tr]{CpMethodCallRange}, if this par block can be verified, it is safe to split this block across endpoints using the endpoint projection. For this rule, the allowed syntaxes for the message and destination are similarly restricted as \refrule[tr]{CpMethodCallRange} to allow for automatic annotation generation.

%% file: appendices/allEpRules.tex
\section{Endpoint Projection Rules}\label{app:ep}

\begin{figure}[ht!]
\centering
\begin{adjustwidth}{-0.2\paperwidth}{-0.2\paperwidth} 
\begin{mathpar}
\RuleEpAssign[tr]{} \and
\RuleEpAssignSkip[tr]{} \and
\RuleEpSend[tr]{} \and
\RuleEpReceive[tr]{} \and
\RuleEpComm[tr]{} \and
\RuleEpCommSkip[tr]{} \and
\RuleEpExpr[tr]{} \and
\RuleEpExprSkip[tr]{} \and
\RuleEpExprIndex[tr]{} \and
\RuleEpRange[tr]{} \and
\RuleEpAnd[tr]{} \and
\RuleEpChor[tr]{} \and
\RuleEpIf[tr]{} \and
\RuleEpWhile[tr]{} \and
\RuleEpIndexSend[tr]{} \and
\RuleEpIndexReceive[tr]{} \and
\RuleEpRangeSend[tr]{} \and
\RuleEpRangeReceive[tr]{}
\end{mathpar}
\end{adjustwidth}
\caption{All endpoint projection rules}
\end{figure}

Rules \refrule[tr]{EpAssign}, \refrule[tr]{EpExpr}, \refrule[tr]{EpAnd}, \refrule[tr]{EpIf}, \refrule[tr]{EpWhile} and their \textsc{*Skip} versions should be self explanatory: they preserve the meaning of the choreographic statement if it is related to the current projection target, and otherwise replace it with the empty block statement (resp. \kwtrue{} for expressions). 

Rule \refrule[tr]{EpComm} shows that each communication statement is processed twice: once in $send$ mode and once in $receive$ mode, passed as an argument through the superscript position. The sending part is processed first to ensure the projected program cannot get stuck. Rule \refrule[tr]{EpCommSkip} shows what happens when the projection target is neither in the sending or the receiving position: it is replaced with the empty block statement. The \textsf{sort} function is used here to avoid having to add duplicate cases for both singular endpoints $e$ as well as endpoint family indices such as \kwfi{}.

Rules \refrule[tr]{EpSend} and \refrule[tr]{EpReceive} shows that sends and receives are encoded with resp. \texttt{writeValue} and \texttt{readValue} method calls on a channel. The specific channel is retrieved from a table that is pre-generated similar to how this is done for the choreographic projection. This is indicated with the $\denote{L}_r$ notation.

Rule \refrule[tr]{EpExprIndex} shows how to transform an endpoint expression when it concerns an indexed endpoint family, and when the current projection target is also an indexed endpoint family. In this case, the expression $E$ is encoded in such a way that it is only evaluated if the indices match of the two endpoint families.

For rule \refrule[tr]{EpRange}, this is similar, except that the current projection target index now has to be in a range $[E_h,)$.

Rule \refrule[tr]{EpChor} always drops the expression $E$, as this annotation indicates the expression should only be included in the choreographic projection.

Rules \refrule[tr]{EpIndexSend}, \refrule[tr]{EpIndexReceive}, \refrule[tr]{EpRangeSend} and \refrule[tr]{EpRangeReceive} apply the same trick as \refrule[tr]{EpExprIndex}, they check if the current projection target index falls in the range specified by the statement. 

For rule \refrule[tr]{EpRangeReceive} there is an additional complication: the inverse of the expression $d$ needs to be computed at run-time to determine the sending endpoint index. While injectivity of this expression is checked using the choreographic projection, this does not result in the actual expression $d^{-1}$; \veymont{} reasons symbolically about it during the choreographic projection, for the purposes of verification. Instead, the endpoint projection uses pattern matching to ensure $d$ is in a form that is actually invertible. For example, the expression $i+1$ can be pattern matched to find that the inverted form is $i-1$.